\documentclass[twocolumn,twocolappendix]{openjournal}

\usepackage{xcolor}
\definecolor{darkgreen}{RGB}{0,100,0}

\usepackage{textgreek}
\usepackage[utf8]{inputenc}
\usepackage[english]{babel}
\usepackage{listings}

\definecolor{codeText}{HTML}{24292F}
\definecolor{codeBlue}{HTML}{0550AE}
\definecolor{codePurple}{HTML}{A626A4}
\definecolor{codeGreen}{HTML}{587C63}
\definecolor{codeTeal}{HTML}{087E8B}
\definecolor{codeBorder}{HTML}{D8DEE4}

\lstdefinestyle{pythonstyle}{
    language=Python,
    basicstyle=\ttfamily\small\color{codeText},
    keywordstyle=\color{codeBlue},
    stringstyle=\color{codePurple},
    commentstyle=\color{codeGreen},
    emph={print,range,len,enumerate,zip,
          isinstance,super,open,sum,min,max},
    emphstyle=\color{codeTeal},
    backgroundcolor=\color{white},
    frame=l,
    rulecolor=\color{codeBorder},
    framerule=1pt,
    framesep=8pt,
    linewidth=\linewidth,
    aboveskip=1em,
    belowskip=1em,
    numbers=none,
    breaklines=true,
    showstringspaces=false,
    keepspaces=true,
    tabsize=4,
    columns=fullflexible,
    upquote=true
}

\usepackage{hyperref}
\hypersetup{
    unicode, 
    colorlinks=true,
    linkcolor=linkcolor,
    citecolor=linkcolor,
    filecolor=linkcolor,
    urlcolor=linkcolor,
}
\usepackage{color,colortbl}
\usepackage{xcolor}
\definecolor{linkcolor}{rgb}{0.0,0.3,0.5}
\usepackage{tensind}
\tensordelimiter{?}
\DeclareGraphicsExtensions{.bmp,.png,.jpg,.pdf}
\usepackage{verbatim}
\usepackage[normalem]{ulem}
\usepackage{orcidlink}
\usepackage{soul}
\graphicspath{ {./figs/} }

\usepackage{tabularx}
\usepackage{listings}
\usepackage{amsmath}

\begin{document}

\title{\textsc{py4radiation}: a framework for generating heating/cooling functions and ion fractions from custom radiation fields for astrophysical simulations}

\author{Daniel Villarruel-Y\'anez\orcidlink{0009-0002-5840-6567}}
\email{daniel.villarruel@yachaytech.edu.ec}
\affiliation{Grupo de Astrof\'isica y Cosmolog\'ia, Escuela de Ciencias F\'isicas y Nanotecnolog\'ia, Universidad Yachay Tech. Hacienda San Jos\'e S/N, 100119, Urcuqu\'i, Ecuador}

\author{Wladimir E. Banda-Barrag\'an\orcidlink{0000-0002-1960-4870}}
\affiliation{Grupo de Astrof\'isica y Cosmolog\'ia, Escuela de Ciencias F\'isicas y Nanotecnolog\'ia, Universidad Yachay Tech. Hacienda San Jos\'e S/N, 100119, Urcuqu\'i, Ecuador}
\affiliation{Hamburger Sternwarte, University of Hamburg, Gojenbergsweg 112, 21029, Hamburg, Germany}

\author{Benedetta Casavecchia\orcidlink{0000-0003-3860-3248}}
\affiliation{Kavli IPMU (WPI), UTIAS, The University of Tokyo, Kashiwa, Chiba 277-8583, Japan}
\affiliation{Max-Planck-Institut für Astrophysik, Karl-Schwarzschild-Strasse 1, 85748, Garching b. München, Germany}

\begin{abstract}
     Modelling multiphase gas in and around galaxies is key to understanding the baryon cycle in galaxy evolution. Modern simulation codes rely on heating/cooling functions and ion fractions to capture the thermodynamics and chemistry of such gases, but coupling all physical factors remains challenging. We present a new Python-based, parallelised package that acts as an integration framework, connecting well-established astrophysical codes and models to produce self-consistent heating/cooling functions and ion fractions from custom ionising UV/X-ray backgrounds. The package returns multi-dimensional tables that can be readily coupled to hydrodynamic and post-processing codes to study radiative processes and extract observables, as we demonstrate with static cloud simulations. Our simulation suite studies the thermodynamic history of a static circumgalactic cloud with smoothed edges subjected to meta-galactic and starburst heating. We find that the clouds experience damped pulsations regulated by strong cooling-driven inflows and heating-driven outflows that constantly overshoot pressure equilibrium. Meta-galactic heating is weak and produces a confined and dense cloud, while starburst heating is strong and sustains large clouds only a little smaller than those obtained with the standard $10^4\,\rm K$ cooling floor. Starburst radiative heating acts as a damping agent, thereby reducing mixing. \textsc{py4radiation} comes with extensive documentation and a sample catalogue of heating/cooling functions and ion tables for a selected set of input meta-galactic and starburst spectral energy distributions. The code is publicly available and can be easily expanded.
\end{abstract}


\maketitle

\section{Introduction}
\label{section:intro}
Studying the role of UV/X-ray radiation fields on interstellar and circumgalactic media is key to understanding the cycle of matter and energy in galaxies \citep{HaardtMadau2012,Tumlinson2017,Faucher-Giguere2023,Wiersma2009}. Feedback processes, either from starbursts (SB) or active galactic nuclei (AGN), result in the formation of large-scale galactic winds, which transport gas, dust, and metals from the interstellar medium (ISM) out to the circumgalactic (CGM) and intergalactic media (IGM) \citep{Veilleux2005, Veilleux2020}. The detailed physics and chemistry of the multi-phase, outflowing gas as well as the metal ion populations are regulated by several processes, such as the mechanical launching of the wind itself via energy and momentum inputs from SB and AGNs \citep{Chevalier1985, Veilleux2005,FaucherGiguere2012}, the mass loading and entrainment efficiencies of cold gas and dust into the hot flow \citep{Gronke2018,Chen2024}, the generation of turbulence \citep{Fielding2020}, the thermal balance regulated by radiative processes (both heating and cooling) \citep{BandaBarragan2021,Faucher-Giguere2023,2025MNRAS.540.3798A}, and the magnetic stresses exerted by global magnetic fields \citep{BandaBarragan2018, HidalgoPineda2024, Casavecchia2024}. Micro-physical processes also play an important role, such as cold-cloud coagulation \citep{Gronke2023}, shattering \citep{Sparre2019}, thermal conduction \citep{Brueggen2016}, self-gravity \citep{Li2020}, and hydrodynamic shielding of dense gas \citep{2019AJ....158..124F,Villares2024}.\par

Among the above, radiative processes are of particular interest as they control the thermodynamic state of the gas, both at the wind launching site in the ISM and at several distances along the CGM and IGM of the host galaxies \citep{Wiersma2009, Oppenheimer2013, DiFederico2026}. Naturally, UV radiation fields are stronger close to their sources (either AGNs or SBs) and weaker farther away, so the conditions of gas in close proximity to the ionising sources are different from those experienced by gas far away from the sources \citep{HaardtMadau2012, Zhu2024}. In this paper, we focus our attention on SB feedback processes regulated by star formation. In such cases, the gas metallicity, the evolutionary stage of the SB (its age), and the initial mass function (IMF) can also play a significant role in determining the shape and strength of the resulting SB UV field \citep{Leitherer1999}. To account for the physical properties of the central SB, its age, and the gas metallicity, it is essential to compute heating and cooling rates as well as ion fractions in a self-consistent manner, i.e. by considering suitable spectral energy distributions (SEDs) of the central SB as well as appropriately calculated flux densities at different distances from the SB source. Such a treatment is relevant not only for accurately modelling the thermodynamic and ionisation structure of multiphase gas, but also for predicting synthetic observables, such as emission- and absorption-line signatures \citep{2021ApJ...919..112D,Casavecchia2024, Villarruel2026}, that can be compared with observations of galactic outflows at different redshifts \citep{2013ApJ...770L...4M,Sugahara2019}.\par

Several astrophysical packages have been developed throughout the years to model the physics and chemistry of gas near ionising sources. In particular, \textsc{Starburst99} (Sb99; \citealt{Leitherer1999}) and \textsc{Cloudy} \citep{Ferland2013} can be used in tandem to produce self-consistent ion fractions and radiative heating and cooling rates. On the one hand, Sb99 has been extensively used to compute the SEDs associated with SB feedback processes. Sb99 takes the IMF, the gas metallicity, and SB conditions (instantaneous or continuous SF models) to calculate the distribution of SB luminosity as a function of wavelength and SB age. On the other hand, the \textsc{Cloudy} code allows users to synthesise the detailed chemistry, ion fractions, and radiative heating and cooling rates of plasmas subjected to different photo-ionising UV fields. Thus, the appropriately normalised and distance-adjusted UV flux densities from Sb99 can be used as inputs for the \textsc{Cloudy} code. Unfortunately, publicly available and user-friendly pipelines to combine both tools are scarce in the literature, with the Perl-based \textsc{CIAOLoop}\footnote{\url{https://github.com/brittonsmith/cloudy_cooling_tools}} code \citep{Smith2008} being a notable exception.\par

Motivated by the above, we present in this paper our new Python-based, parallelised integration framework, \textsc{py4radiation}, that tackles these limitations by providing a user-friendly interface to link any input meta-galactic or starburst UV background (e.g. an SB SED from the Sb99 database) with the \textsc{Cloudy} code via a modernised Python implementation of \textsc{CIAOLoop}. \textsc{py4radiation} is publicly available and released under the MIT licence. It aims to facilitate the generation of heating and cooling coefficients together with ion fraction tables, consistent with input meta-galactic UV/X-ray or SB UV radiation fields. The details of the Python code are presented in section \ref{section:code-implementation}, usage examples as well as sample radiative heating/cooling coefficients and ion fractions can be found in sections \ref{section:code-execution-performance} and \ref{section:application-demo}. Lastly, we discuss the limitations and future code development in section \ref{section:conclusions}. Jointly with the code, we also release a sample catalogue of tabulated heating/cooling rates and ion fractions for a sample set of meta-galactic and SB conditions, which can be directly downloaded by users.

\section{Code Implementation and Documentation}
\label{section:code-implementation}

\begin{figure}
    \centering
    \includegraphics[width=\linewidth]{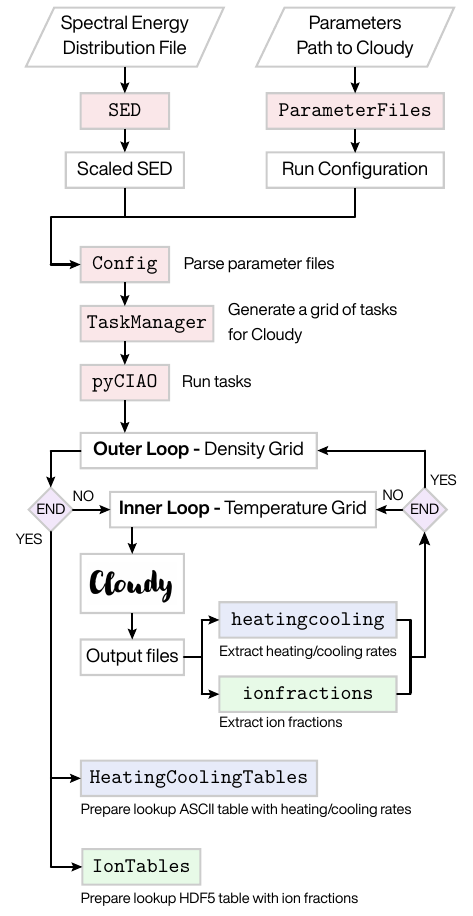}
    \caption{Workflow of \textsc{py4radiation}. The inner and outer loops produce several \texttt{Cloudy} simulations, one for each point in the defined temperature-number density grid. Depending on the selected mode, \textsc{py4radiation} produces radiative heating and cooling rates (blue undertone) or ion fractions (green undertone).}
    \label{fig:workflow}
\end{figure}

\textsc{py4radiation} is a Python-based, object-oriented, modular, and parallelised package designed to prepare custom spectral energy distributions for photo-ionisation calculations and construct lookup tables of radiative heating, cooling, and ionisation fractions. The resulting outputs can be readily coupled to hydrodynamical simulation codes to modify the internal energy of the system with radiative gains and losses, as well as with analysis, post-processing tools to produce synthetic observables (e.g., ion column densities and emission/absorption spectra of selected ion species). The package is organised into three principal components: (i) preparation of the incident radiation field and configuration of the physical parameter space, (ii) numerical execution of \textsc{Cloudy} calculations through \texttt{pyCIAO}, and (iii) post-processing and construction of lookup tables. The complete workflow is illustrated in Fig. \ref{fig:workflow}.\par

The current implementation supports the meta-galactic UV/X-ray background of \cite{HaardtMadau2012}, pre-computed stellar population spectra from \textsc{Starburst99} \citep{Leitherer1999}, and user-provided SEDs. The photo-ionisation calculations are performed using version C13.05 of \textsc{Cloudy} \citep{Ferland2013}.\par

The software, its source code, installation instructions, the Application Programming Interface (API) reference, and worked examples are publicly available through the project repository and documentation\footnote{Repository: \url{https://github.com/cPhysPlus/py4radiation}\\Documentation: \url{https://py4radiation.readthedocs.io}}. The tutorials include representative numerical outputs that can be used to inspect the calculations and compare results obtained with independent installations.

\subsection{Preparation of radiation fields}
\label{subsection:radiation-fields}

The first stage of the workflow consists of selecting an incident radiation spectrum and defining the physical parameter space over which the calculations are performed. The \texttt{SED} class retrieves or loads the selected spectrum, applies the appropriate physical scaling, and prepares a \textsc{Cloudy}-readable input file.\par

For stellar radiation, \textsc{py4radiation} supports the original \textsc{Starburst99} models presented by \cite{Leitherer1999}. Following their recommendations for photo-ionisation calculations, the package uses the stellar-only continua models. The supported models include both instantaneous burst and continuous star formation. For each model, the user selects a stellar population age from the tabulated values. The reference luminosities can be rescaled according to the star formation rate (SFR) for continuous models or the initial stellar mass for instantaneous bursts. The corresponding luminosity scaling is:
\begin{equation}
    L_{\lambda} = L_{\lambda,0}
    \begin{cases}
        \dfrac{\rm SFR}{1\,M_{\odot}\,{\rm yr}^{-1}}, & \text{continuous star formation}, \\[10pt]
        \dfrac{M_{\star}}{10^6\,M_{\odot}}, & \text{instantaneous burst},
    \end{cases}
\end{equation}
where $L_{\lambda,0}$ is the spectral luminosity of the reference Sb99 model. For stellar sources, the spectral luminosity is converted into a specific flux density assuming isotropic emission and geometrical dilution.

The package also supports the meta-galactic UV/X-ray background presented by \citet{HaardtMadau2012} (H\&M2012), currently set to redshift $z=0$ only. Users seeking other redshifts can easily add the $z$ identifier to the parameter files. For consistency with the spectral representation used for stellar radiation, \textsc{py4radiation} converts the specific intensity $J_{\nu}$ expressed in ${\rm erg}\,{\rm s}^{-1}\,{\rm cm}^{-2}\,{\rm Hz}^{-1}\,{\rm sr}^{-1}$ to angle-integrated intensity, $F_{\nu}$, expressed in ${\rm erg}\,{\rm s}^{-1}\,{\rm cm}^{-2}\,{\rm Hz}^{-1}$. This quantity is used to normalise the incident continuum in \textsc{Cloudy}.\par

Both Sb99 and H\&M2012 tables can be retrieved from their original repositories\footnote{Sb99: \url{https://massivestars.stsci.edu/starburst99}\\H\&M2012: \url{https://www.ucolick.org/~pmadau/CUBA}} and stored in a local cache. We provide retrieval scripts to download these tables inside the \texttt{/utils} sub-package. Alternatively, users can provide previously downloaded tables or their own user-defined spectra. \textsc{py4radiation} supports custom input spectra as two-column tables containing wavelength and logarithmic spectral luminosity. After the appropriate conversion, photon energies are expressed in Rydbergs (Ryd). The resulting files include the spectral shape and its normalisation at 1 Ryd, corresponding to the hydrogen ionisation threshold.

The \texttt{ParameterFiles} class generates the parameter files required to execute the calculations. These parameter files specify the \texttt{Cloudy} executable, the incident radiation field, the physical grid, and the numerical settings of each execution mode. The physical grid considers hydrogen number density ($n_{\rm H}$) and temperature sampled over logarithmic grids, with user-defined limits and resolutions. There are two supported execution modes: (i) generation of heating and cooling maps, and (ii) calculation of ion fractions for selected chemical elements (atomic species up to zinc).

\subsection{Calculation with \texttt{pyCIAO}}
\label{subsection:pyCIAO}

The numerical execution is managed by \texttt{pyCIAO}, a modernised Python-based implementation of the functionality originally implemented by the Perl code \textsc{CIAOLoop}. \texttt{pyCIAO} separates configuration parsing, task generation, code execution, and output management into independent modules, which facilitates both maintainability and parallel execution.\par

The \texttt{Config} class parses the parameter files and prepares sets of commands common to all \textsc{Cloudy} calculations and varying physical parameters, which define the grid for execution. These commands are processed by the \texttt{TaskManager} class, which constructs the Cartesian product of the specified grid values and assigns a unique identifier to each resulting task. Each task represents an independent Cloudy configuration with the required physical parameters and execution commands.\par

The execution of these tasks supports the native \texttt{multiprocessing} module and also distributed-memory parallelisation through the Message Passing Interface (\texttt{MPI}) library, implemented using \texttt{mpi4py} \citep{Dalcin2021}. In \texttt{multiprocessing} mode, tasks are distributed among a pool of local worker processes. Under \texttt{MPI}, the complete set of tasks is distributed among the available ranks. Each rank independently executes the assigned calculations. For each task, \texttt{pyCIAO} evaluates an internal temperature range. Each pair $(n_{\rm H},\,T)$ corresponds to an individual \textsc{Cloudy} execution. The resulting output files are processed according to the selected calculation mode.\par

The selected execution mode determines the files and quantities that are processed by \texttt{pyCIAO}. The results are organised into files associated with their corresponding task identifiers. \texttt{pyCIAO} also prepares a master file (a user-defined name with a \texttt{.run} extension) to record the combination of parameters, which is required to reconstruct the numerical grid utilised during post-processing.

\subsection{Construction of tables}
\label{subsection:construction-tables}

The final stage of the workflow prepares structured, multi-dimensional lookup tables from \texttt{pyCIAO} outputs. The classes \texttt{HeatingCoolingTables} and \texttt{IonTables} are responsible for preparing the tables for heating/cooling coefficients and ionisation fractions, respectively. Both classes use the hydrogen number density recorded in the master run file as the outer loop and the temperature as the inner loop to create the tables with the calculated value for each point of the grid.\par

The volumetric heating and cooling coefficients are normalised by $n_{\rm H}^2$ or $n_{\rm H}\,n_{\rm e}$, where $n_{\rm e}$ is the electron number density. The normalisation is chosen by the user. The units of the heating and cooling coefficients are expressed in ${\rm erg}\,{\rm cm}^3\,{\rm s}^{-1}$. Also, the mean particle mass, $\mu$, is calculated using elemental abundances and ion populations extracted from \textsc{Cloudy} outputs. The \texttt{HeatingCoolingTables} class produces a single \texttt{ASCII} table containing $(n_{\rm H},\,T,\,{\rm heating},\,{\rm cooling},\,\mu)$. This dataset provides the thermodynamic quantities required to incorporate density- and temperature-dependent radiative source terms into hydrodynamical and magnetohydrodynamical simulations.\par

For the ionisation tables, given a chemical element, \texttt{pyCIAO} extracts the logarithmic ionisation fraction for each ion state and gathers them into multi-dimensional arrays. The resulting ion tables are stored in the Hierarchical Data Format version 5 (\texttt{HDF5}, \citealt{Folk2011}), with an individual dataset for each chemical element. The \texttt{HDF5} structure is designed to be compatible with the ionisation table format utilised by \textsc{Trident} \citep{Hummels2017} and \textsc{yt} \citep{Turk2011}. For our redshift-zero calculations, the ion table generation includes a duplicated redshift coordinate to accommodate the multi-dimensional interpolation requirement of these tools. Users interested in studying the cosmological redshift evolution of the lookup tables can do so by performing an additional loop over a desired redshift range.

\section{Code Execution and Performance}
\label{section:code-execution-performance}

\subsection{How to use \textsc{py4radiation}}
\label{subsection:how-to-use}

In this section we show two code insets: In the first one, we provide an annotated example Python script for obtaining the parameter files to generate heating/cooling rates and ion fractions from a sample Sb99 SED (fig8b from \citealt{Leitherer1999}). This is an SED with stellar-emission only, corresponding to a $3$-Myr-old stellar population, with an SFR of $1$ $M_{\odot}$ yr$^{-1}$, a Salpeter initial mass function (IMF) with slope $\alpha=2.35$ (with $M_{\rm low}=1$ $M_{\odot}$ and $M_{\rm up}=100$ $M_{\odot}$), and scaled to $5$ kpc from the radiation source. In the second one, we illustrate how to obtain the lookup tables from the intermediate outputs.\par

The first code inset below shows how to load the relevant classes from \textsc{py4radiation}, set the name of the model, load one of the downloaded Sb99 SEDs with its identifier according to \cite{Leitherer1999}, and indicate the specific distance to the radiation source. The spectrum file needed for execution is readily generated and formatted to be \textsc{Cloudy}-readable by \texttt{pyCIAO}. The parameter files require specifying the path to the \textsc{Cloudy} executable, the model name, and the desired elements using chemical symbols in a single string. In the code inset, we include hydrogen, carbon, nitrogen, and oxygen. We also define the logarithmic parameter grid limits (for number density and temperature) and the number of points in each dimension. At the end, we generate parameter files for both execution modes.\par

\begin{lstlisting}[label={code:inset1}]
from py4radiation import SED, ParameterFiles

model = "sb5"
elements = "H C N O"

sed = SED(
    run_name=model,
    table_id="8b",
    age_myr=3,
    sfr_msun_yr=1.0,
    distance_kpc=5.0)
    
sed.write_outfile()

par = ParameterFiles(
    cloudy_path="path/to/cloudy.exe",
    run_name=model,
    elements=elements,
    normalisation=1, # (1) n_H^2; (2) n_e * n_H
    n_limits=(-6, 4),
    T_limits=(2, 8),
    n_spacing=None,
    n_points=256,
    T_points=256)

par.hc_parfile()
par.ib_parfile()
\end{lstlisting}

The resulting parameter files, named \texttt{sb5\_hc.par} and \texttt{sb\_ib.par}, define the heating/cooling and ionisation calculations to be performed by \textsc{Cloudy}, respectively. \texttt{pyCIAO} can be executed from the command line either with local processes via \texttt{multiprocessing} or using distributed-memory execution launched through \texttt{MPI}. The number of \texttt{MPI} processes is specified through the launcher, and \texttt{pyCIAO} distributes the independent tasks among the available or specified ranks. The reader is referred to the code documentation for a detailed tutorial.\par

The \textsc{Cloudy} commands in the \texttt{.par} files specify gas in stationary ionisation balance under the radiation field specified by the user, and consider single-zone, optically-thin calculations within the prescribed ranges for hydrogen number densities and temperatures. By default, \textsc{Cloudy} assumes a Solar abundance mixture \citep{Ferland2013}. Dust grains, cosmic rays, and the cosmological background are not included. Nevertheless, the user may modify the parameter files using the \texttt{command} keyword to include any additional \textsc{Cloudy}-specific setting.\par

During execution, \texttt{pyCIAO} writes the \textsc{Cloudy} simulation results of individual calculations into the directories specified in the parameter files. After all \textsc{Cloudy} runs have finished, the outputs need to be gathered and compiled into lookup tables. The second code inset below shows how to load the relevant output functions, define the directories where the \texttt{pyCIAO} outputs are located, provide the model name, and, in the case of ion tables, include the elements that were calculated. At the end, both tables are written.\par

\begin{lstlisting}
from py4radiation import IonTables, HeatingCoolingTables

ib = IonTables(
        run_dir="ib",
        run_name=model,
        elements=elements)

hc = HeatingCoolingTables(
        run_dir="hc",
        run_name=model)

ib.write_table()
hc.write_table()
\end{lstlisting}

The resulting tables for each mode, \texttt{sb5\_iontable.h5} and \texttt{sb5\_cooltable.dat}, are saved in the local working directory. The procedure is independent of the input SED, provided that the incident spectrum can be expressed in the input format supported by \textsc{py4radiation}. The same workflow can be applied to different stellar population models, different source distances, or externally supplied radiation fields without modifying the underlying computational routines.

\subsection{Sample Outputs from \textsc{py4radiation}}
\label{subsection:sample-output}

\begin{figure}
    \centering
    \includegraphics[width=\linewidth]{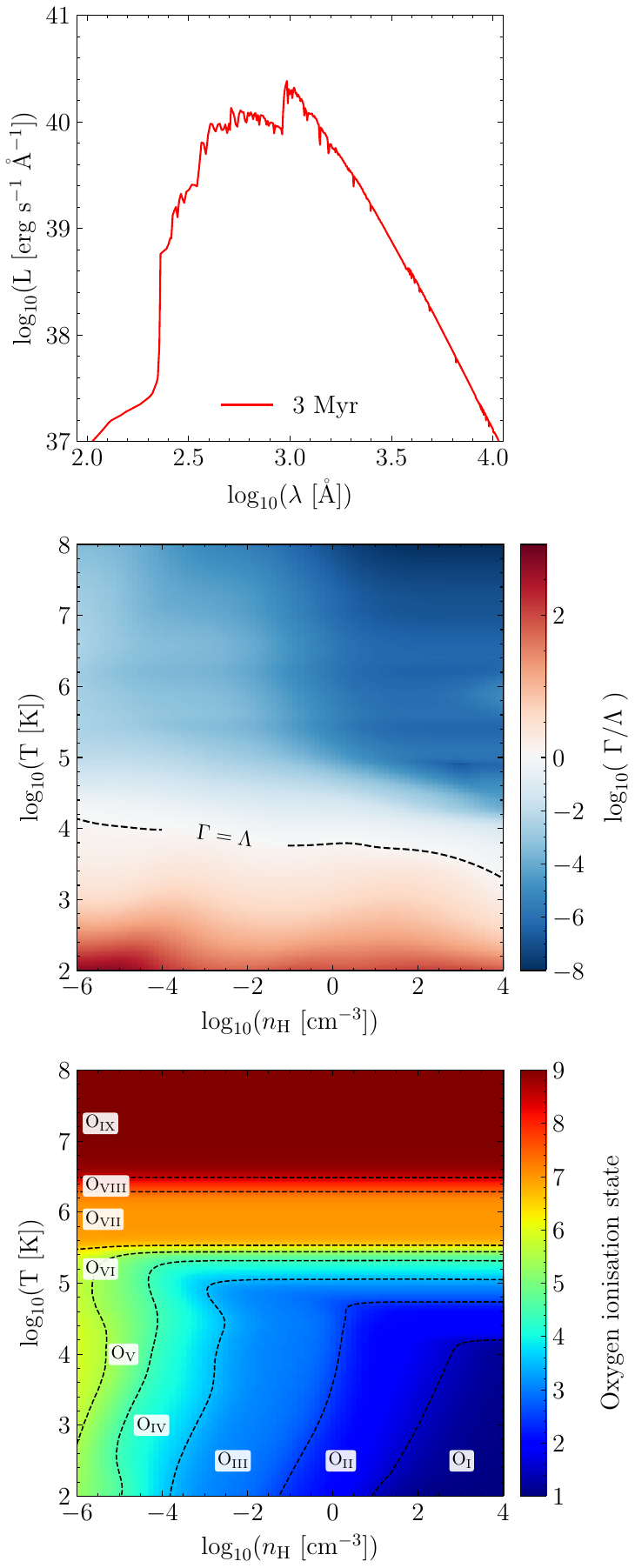}
    \caption{Input SED (top panel) and sample outputs (middle and bottom panels) of the annotated example presented in subsection \ref{subsection:how-to-use}. \textsc{py4radiation} takes the spectral energy distribution and calculates heating/cooling rates and ion fractions across the defined parameter grid, $(n_{\rm H},T)$, and the specified physical conditions. The resulting tables are presented as the heating-to-cooling ratio (middle panel) and the ionisation-state diagram for O {\sc vi} (bottom panel) across $T-n_{\rm H}$ phase space.}
    \label{fig:sample-output}
\end{figure}

Figure \ref{fig:sample-output} shows the input SED and sample outputs obtained from the annotated example described in the previous section. The upper panel shows the Sb99 spectrum, which \texttt{pyCIAO} uses to compute heating/cooling rates and ion fractions over the prescribed grid of hydrogen number densities and temperatures. The middle panel presents the logarithm of the ratio between heating $\Gamma$ and cooling $\Lambda$ rates, calculated for the input SED. Positive (red) values indicate regions where heating dominates over cooling, and negative (blue) values correspond to cooling-dominated regions. The dashed line shows where thermal equilibrium sits for the adopted radiation field. Appendix \ref{ap:comparison} shows results consistent with other heating/cooling curves in the literature. The bottom panel shows the dominant ionisation state of oxygen at each number density and temperature. These sample outputs demonstrate how the resulting lookup tables characterise both the thermal and ionisation properties of gas exposed to a specified radiation field, providing the quantities required for use in numerical simulations and post-processing routines.\par

\subsection{Parallel Performance}
\label{subsection:parallel}

The parallel performance of \texttt{pyCIAO} was evaluated with \textsc{Cloudy} C13.05 on the SuperMUC-NG supercomputer at the Leibniz Supercomputing Centre (LRZ), Germany. We conducted strong-scaling tests using the heating and cooling mode, with a fixed computational workload distributed across an increasing number of \texttt{MPI} processes. The logarithmic parameter grid comprised $256$ hydrogen number densities and $256$ temperatures, corresponding to a total of $65,536$ individual \textsc{Cloudy} executions. We performed this benchmark using $16$, $32$, $64$, $128$, and $256$ \texttt{MPI} processes distributed across one to eight compute nodes in the \texttt{micro} partition. The first two configurations utilised $16$ and $32$ processes on a single node, while the remaining configurations used $32$ processes per node.\par

\begin{figure}
    \centering
    \includegraphics[width=0.92\linewidth]{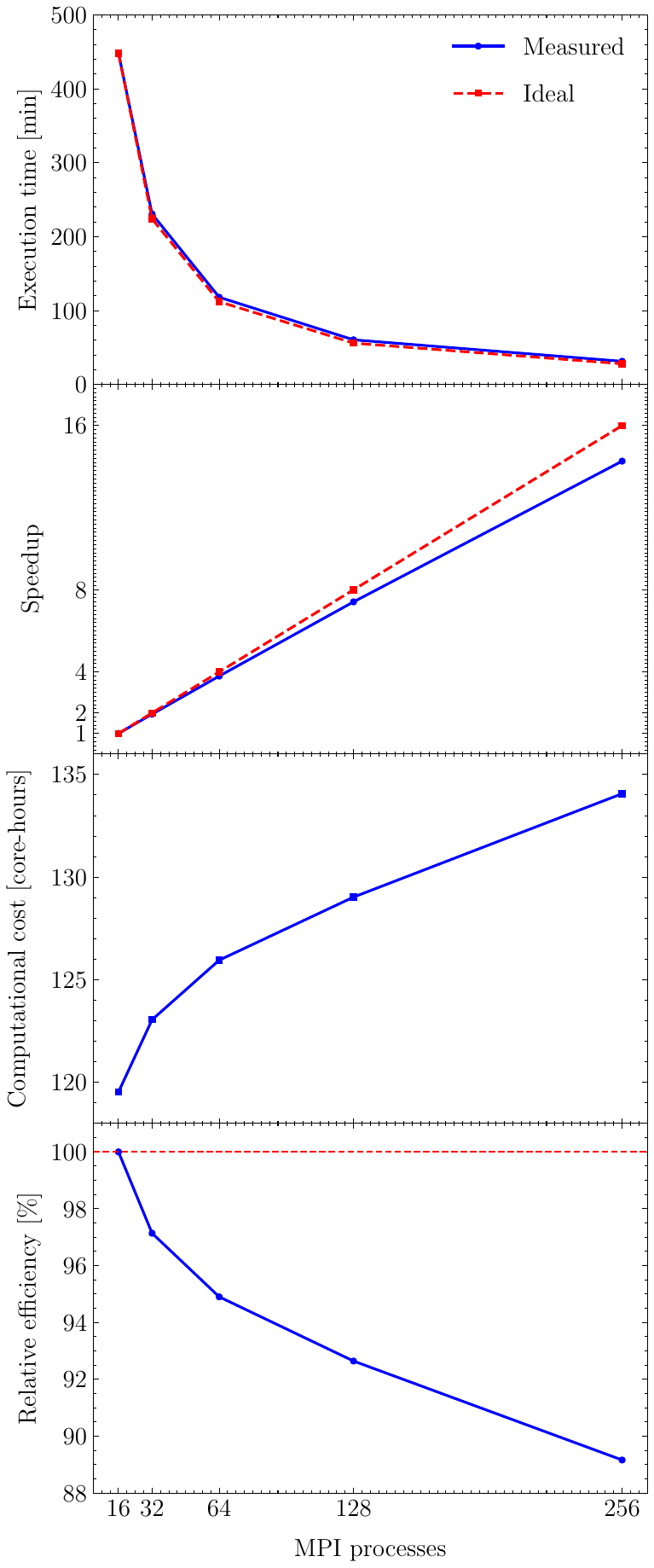}
    \caption{Strong-scaling performance of \texttt{pyCIAO} as a function of the number of \texttt{MPI} processes. From top to bottom, the panels show the execution time, the speedup, the computational cost (in core-hours), and the relative parallel efficiency. The speedup and relative efficiency are calculated using 16 \texttt{MPI} processes as the reference configuration.}
    \label{fig:scaling}
\end{figure}

The top two panels of Fig. \ref{fig:scaling} show the measured execution times and parallel speedup as functions of the number of \texttt{MPI} processes. The execution time decreases from $\sim 450$ minutes using 16 processes to $\sim 30$ minutes using 256 processes. The parallel speedup is defined relative to the smallest configuration. At $256$ processes, \texttt{pyCIAO} achieves a speedup $\gtrsim 14$, compared with the ideal value of $16$. The bottom two panels of Figure \ref{fig:scaling} show that the computational cost increases from $\sim 120$ core-hours at $16$ processes to $\sim 130$ core-hours at $256$ processes. At $256$ processes, the parallel efficiency relative to the smallest configuration shows that \texttt{pyCIAO} achieves $\sim 90\%$.\par

These results are consistent with the task-based parallelisation strategy implemented in \texttt{pyCIAO}. Since each point in the parameter grid is evaluated independently, \texttt{MPI} processes do not require inter-process communication during execution, except for scattering and gathering. At $256$ \texttt{MPI} processes, each rank is assigned a single density-grid task, corresponding to $256$ temperature calculations. The number of points in the logarithmic density grid determines the maximum number of concurrently executable tasks.

\section{Application Demonstration: UV-heated Circumgalactic Clouds}
\label{section:application-demo}

\begin{figure}
    \centering
    \includegraphics[width=0.92\linewidth]{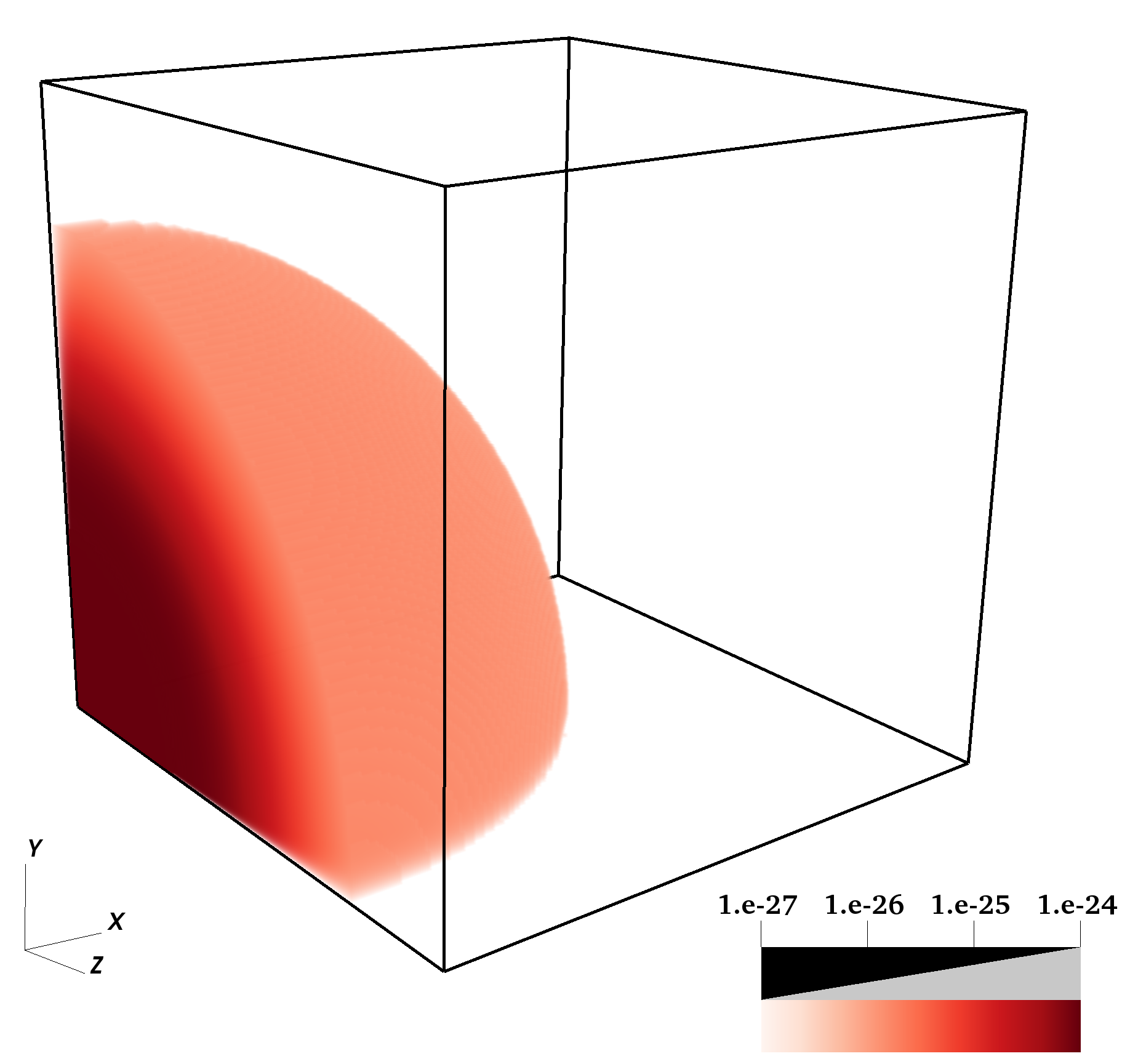}
    \caption{3D rendering showing the simulation domain with a CGM cloud embedded in a more diffuse ambient halo gas. The colour bar shows the mass density field on a logarithmic scale and units of $\rm g\,cm^{-3}$, combined with a linear opacity ramp decreasing from low to high density values. Only 1/8th of the volume is simulated to optimise computing resources.}
    \label{figure:3d_domain}
\end{figure}

To illustrate how the code can be used in conjunction with hydrodynamical packages, we utilise \textsc{py4radiation} to study how starburst UV heating influences the evolution of idealised, static clouds in the CGM. Each 3D CGM cloud is embedded in a much more diffuse ambient medium with halo properties (see Figure \ref{figure:3d_domain}) and is subjected to radiative heating/cooling by different UV/X-ray backgrounds. We use the \textsc{gPLUTO} code (version 0.91; \citealt{Mignone2007, Rossazza2026}) to model the hydrodynamics, and \textsc{py4radiation} to obtain heating and cooling functions consistent with the target radiation background.

\subsection{Spectral Energy Distributions (SEDs)}
\label{subsection:SEDs}
The thermodynamic state of an astrophysical gas depends, among other factors, on the spectral shape and the intensity of the incident radiation field. A common simplification used in numerical simulations of CGM systems as well as in analysis tools (e.g., \citealt{Casavecchia2024,2025MNRAS.540.3798A,Hummels2017}) involves utilising a meta-galactic background (e.g., \citealt{HaardtMadau2012}). While such a background captures the volume-filling, cosmic UV/X-ray radiation field, it does not accurately model the properties of radiation fields near star-forming regions (see \citealt{Villarruel2026}).\par

\begin{figure}
    \centering
    \includegraphics[width=0.92\linewidth]{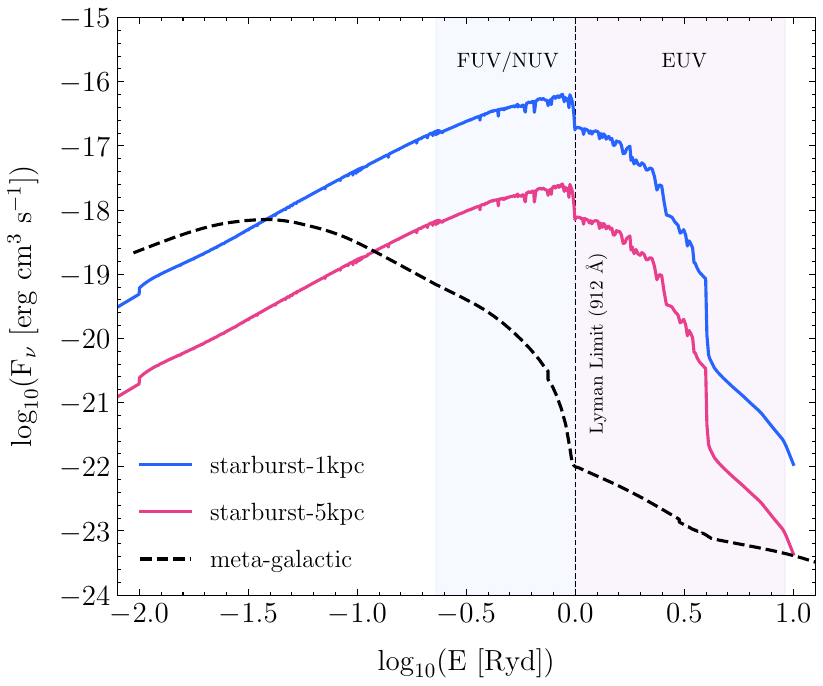}
    \caption{Comparison between the Sb99 SED presented in figure 8b of \cite{Leitherer1999}, at 1 kpc (starburst-1kpc, blue) and 5 kpc (starburst-5kpc, pink) from the radiation source and the \cite{HaardtMadau2012} UV/X-ray cosmic background (meta-galactic, black). There is a roughly constant 2-dex difference between starburst-1kpc and starburst-5kpc, and a 4-dex difference between the latter and the meta-galactic background.\\}
    \label{fig:sed_comparison}
\end{figure}

Figure \ref{fig:sed_comparison} shows a comparison between the \cite{HaardtMadau2012} UV/X-ray cosmic background (meta-galactic) and the distance-scaled SED from figure 8b of \cite{Leitherer1999} at 1 and 5 kpc (starburst-1kpc and starburst-5kpc, respectively). This Sb99 SED assumes gas at those custom distances from a 3-Myr old starburst. The simple inverse-square law scaling preserves the field shape while reducing the flux density by 2 dex. In the CGM surrounding starburst galaxies, the radiation field is dominated by the collective input of massive OB stars. Thus, the Sb99 SEDs show significantly stronger fields at the Lyman limit in the UV region of the spectrum. In fact, at the Lyman limit, the Sb99 SED at 5 kpc shows a 4-dex difference in specific flux density compared to the meta-galactic one.

\subsection{Radiative Heating and Cooling Rates}
\label{subsection:radiative_heating_cooling}

Next, we run \textsc{py4radiation} with a parameter grid of $(n_{\rm H},\,T)=(105,\,321)$ points providing the SEDs described above as inputs (see Figure \ref{fig:sed_comparison}). The resulting heating ($\Gamma$; red) and cooling ($\Lambda$; blue) rates in units of $\rm erg\,cm^3\,s^{-1}$, are shown in Figure \ref{figure:heat_cool}. The lines show the energy rates for gas with different temperatures at three distinct number densities, namely $1$, $0.01$, and $0.001$ ${\rm cm}^{-3}$, which reflect the conditions set for the CGM cloud, expected for mixed gas, and set for the ambient halo, respectively. The background UV radiation field has significant effects on both heating and cooling rates:

\begin{figure*}
    \centering
    \includegraphics[width=\linewidth]{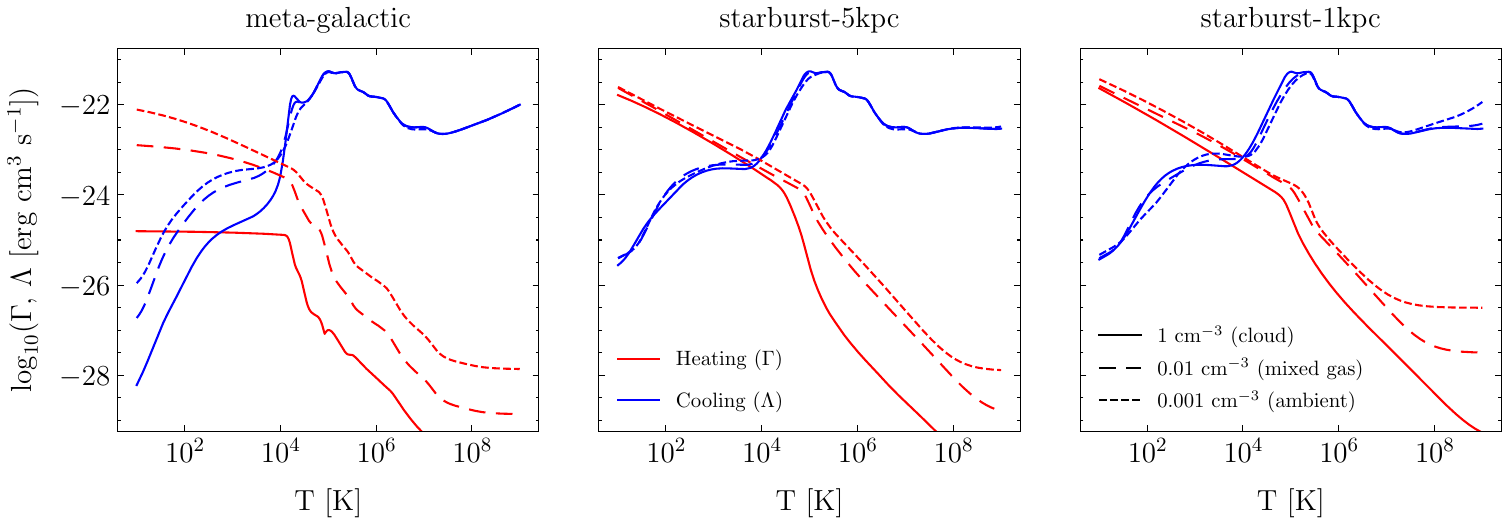}
    \caption{Heating and cooling coefficients as functions of temperature for three values of number density, $n=1\,{\rm cm}^{-3}$ corresponding to the CGM cloud, $n=0.01\,{\rm cm}^{-3}$ corresponding to mixed gas, and $n=0.001\,{\rm cm}^{-3}$ corresponding to the ambient medium. While the meta-galactic field mainly influences the heating/cooling rates of diffuse gas, both starburst fields have a significant impact on the shape and rates of the heating/cooling curves of gas at different densities.}
    \label{figure:heat_cool}
\end{figure*}

\begin{itemize}
    \item \textbf{Effects on heating rates}: At $T<10^4\,\rm K$, there are sharp differences between the meta-galactic and starburst heating rates. While the weak H\&M2012 background is only efficient at photo-heating diffuse gas, the strong starburst backgrounds can effectively heat up both diffuse and dense gas. Heating rates remain high and within less than a dex. for all the densities. At high temperatures, heating becomes inefficient in all cases due to the lack of UV absorption by the fully ionised gas. While the starburst backgrounds allow for slightly higher photo-heating than the meta-galactic background, all backgrounds experience a drastic drop in the heating rates for $T>10^4\,\rm K$.
    \item \textbf{Effects on cooling rates}: At low temperatures $T<10^4\,\rm K$, starburst radiation fields produce slightly higher cooling rates than the meta-galactic field. Being weaker the H\&M2012 background has very little effect on dense gas, while the starburst cooling rates are similar for all densities. At high temperatures, the overall shape of the functions is different, with the two starburst curves showing no Lyman alpha peak. While the weaker meta-galactic background allows for HI to exist and be collisionally excited, the strong UV field of starburst backgrounds make neutral hydrogen virtually absent. Another minor difference can be seen at $T>10^7\,\rm K$, where free-free emission is enhanced in the intense starburst-1kpc model owing to more efficient Compton scattering.
\end{itemize}

We note that the approach commonly used in the literature of setting a cooling floor at $\sim 10^4\,\rm K$ would not fully capture the detailed thermodynamics of gases associated with distinct heating sources. Thus, \textsc{py4radiation} can aid in constructing more thermodynamically- and chemically-consistent numerical models as we demonstrate in the sections below.

\subsection{Simulation set-up: CGM clouds}
\label{section:simulation}
To model the effects of different radiation fields on CGM clouds in a self-consistent manner, we couple the equations of hydrodynamics with the tabulated heating and cooling functions obtained from \textsc{py4radiation}. To do this, we implement a bilinear interpolation method into the \textsc{gPLUTO} code that reads in the heating and cooling rates (in logarithmic space) from a structured $(n_{\rm H},\,T)$ mesh and uses them to add and remove energy from the energy conservation equation. Our simulation set-up comprises a 3D domain with size $20\times 20\times 20\,\rm pc^3$, which represents $1/8$th of the volume occupied by a CGM cloud inside a hot ambient medium (see Figure \ref{figure:3d_domain}). This domain choice ensures an optimal use of computational resources, and adequately captures the evolution of spherically-symmetric, radiative CGM clouds. The computational domain includes a CGM cloud with a temperature of $10^4\,\rm K$ embedded a in hot, diffuse ambient medium at $10^7\,\rm K$. The cloud, with radius of $10\,\rm pc$, has smoothed edges, so its density field smoothly drops from a dense core with $n=1\,\rm cm^{-3}$ to the more diffuse ambient gas with $n=1\times 10^{-3}\,\rm cm^{-3}$, following \cite{BandaBarragan2016}. Thus, the density contrast between the cloud core and the ambient medium is $10^3$. Both the cloud and ambient gas are initially in pressure equilibrium with $P/k_b = 10^{4}\,\rm K\, cm^{-3}$, so that in all our models radiative processes can act on the gas from the very beginning of the simulation at $t=0$.\par

To ensure the correct symmetries are captured, we set up reflective boundary conditions on the $X=0$, $Y=0$, and $Z=0$ planes and outflow boundary conditions for the remaining three sides away from the CGM cloud. To resolve important length scales, we utilise a uniform spatial grid with $128^3$ grid cells, corresponding to a resolution of $32$ cells per cloud core radius. This is an adequate numerical resolution to capture the effects associated with dynamical instabilities as demonstrated by \cite{BandaBarragan2018}.

\subsection{CGM Cloud Models}
\label{subsection:models}

We consider $4$ CGM cloud models with identical initial conditions (as described in the previous section) and varying radiative heating and cooling implementations. First, we consider a standard-cooling model, which includes cooling consistent with the H\&M2012 background, neglects all heating effects, and places a cooling floor at $10^4\,\rm K$. This approach is widely used in the literature to mimic the effects of radiative heating below that temperature. Then, we investigate the impact of three backgrounds: meta-galactic, which assumes the cloud is solely subjected to a meta-galactic UV/X-ray radiation field and there is no starburst contribution; and starburst-5kpc and starburst-1kpc, which consider the CGM cloud is at distances of $5\,\rm kpc$ and  $1\,\rm kpc$, respectively, from a central starburst radiation source. The latter three models have a cooling floor at $10^2\,\rm K$, so that the effects of UV heating can be realised below the usual $10^4\,\rm K$ temperature floor. Table \ref{table:table2} summarises our models, their initial conditions, and the heating/cooling functions employed.\par

\begingroup
\setlength{\tabcolsep}{10pt}
\renewcommand{\arraystretch}{1.3}
\setlength\extrarowheight{2pt}
\begin{table*}[ht!]
    \centering
    \begin{tabularx}{\textwidth}{c c c c c c c c c}
        Model &
        $\rho_{\rm halo}$ &
        $n_{\rm halo}$  &
        $T_{\rm halo}$ &
        $\rho_{\rm cloud}$ &
        $n_{\rm cloud}$  &
        $T_{\rm cloud}$ &
        Heating \& Cooling &
        Cooling Floor \\
         &
        $\rm [g\,cm^{-3}]$ &
        $\rm [cm^{-3}]$   &
        $\rm [K]$  &
        $\rm [g\,cm^{-3}]$ &
        $\rm [cm^{-3}]$   &
        $\rm [K]$  &
         &
         \\
        \hline \hline
        standard-cooling &
        $10^{-27}$ &
        $10^{-3}$  &
        $10^7$ &
        $10^{-24}$ &
        $1$  &
        $10^4$ &
        Meta-galactic &
        $10^4$ \\
        meta-galactic &
        $10^{-27}$ &
        $10^{-3}$  &
        $10^7$ &
        $10^{-24}$ &
        $1$  &
        $10^4$ &
        Meta-galactic &
        $10^2$ \\
        starburst-5kpc &
        $10^{-27}$ &
        $10^{-3}$  &
        $10^7$ &
        $10^{-24}$ &
        $1$  &
        $10^4$ &
        Starburst Heating &
        $10^2$ \\
        starburst-1kpc &
        $10^{-27}$ &
        $10^{-3}$  &
        $10^7$ &
        $10^{-24}$ &
        $1$  &
        $10^4$ &
        Starburst Heating &
        $10^2$ \\
        \hline\hline
    \end{tabularx}
    \caption{Models and initial conditions for our CGM cloud simulations.}
    \label{table:table2}
\end{table*}
\endgroup

\subsection{Simulation results}
\label{subsec:results}

In order to capture a representative sample of the full thermodynamic history of UV-heated CGM clouds, all our simulations run for $10\,\rm Myr$. Figure \ref{figure:density_maps} shows the time evolution of the mass density field in our 4 models (in columns). We show 2D slices at $Z=0$ (on the $XY$ plane) from $t=0$ (top) to $t=2\,\rm Myr$ (bottom). We do not show all the simulated times (up to $t=10\,\rm Myr$ ) as we find most of the dynamical diagnostics reach a damped pulsating steady state after $t=2\,\rm Myr$. In general, the evolution of all models can be summarised in $4$ stages:

\begin{figure*}
    \centering
    \includegraphics[width=\linewidth]{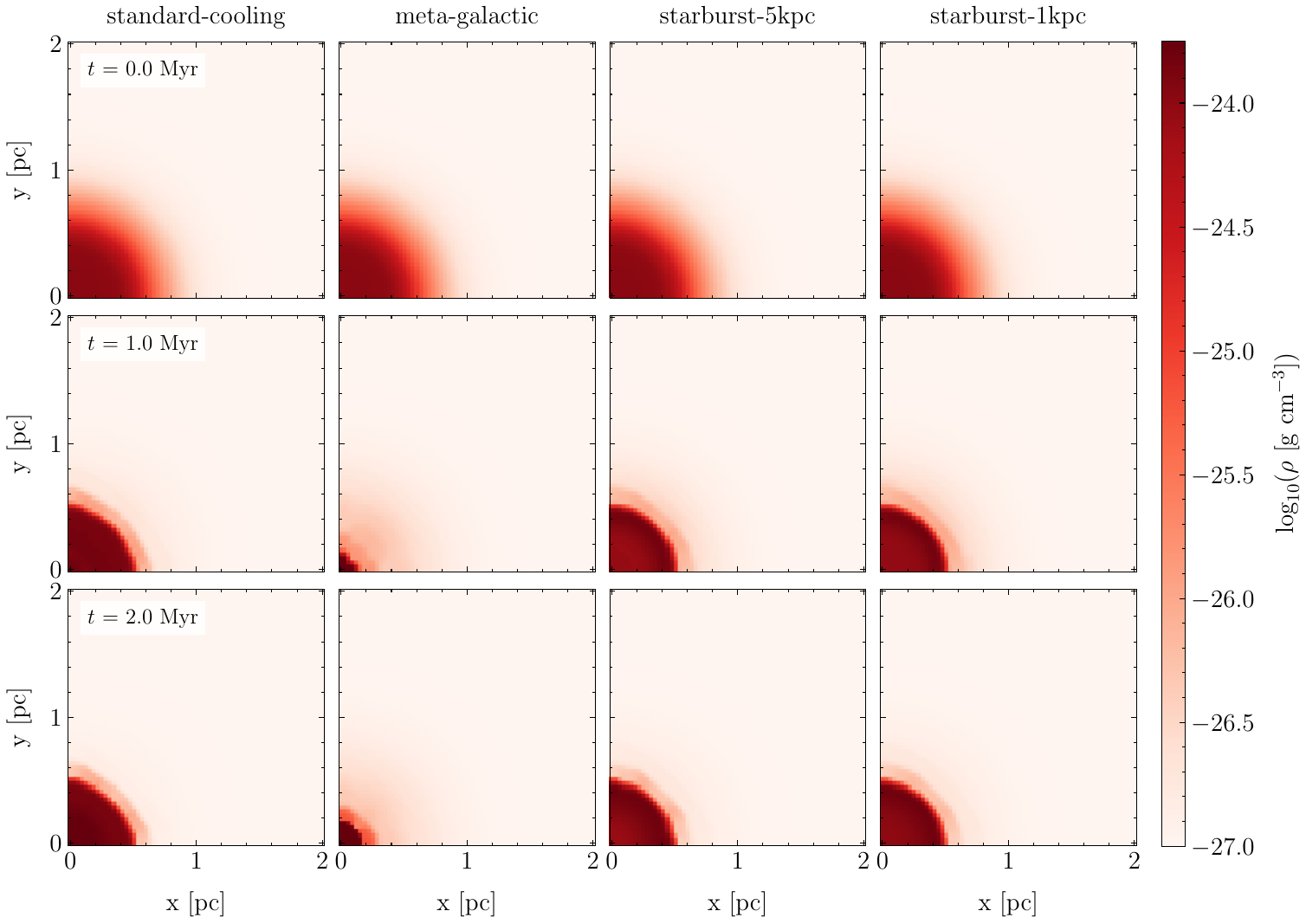}
    \caption{2D slices of the mass density field on the $Z=0$ plane. The standard-cooling and starburst models maintain larger clouds, while the meta-galactic model undergoes substantial cloud compression. All clouds experience acoustic pulsations owing to cooling-driven inflows and heating-driven outflows. A movie showing the full evolution of the four simulations of this work is available in the code repository.}
    \label{figure:density_maps}
\end{figure*}

\begin{enumerate}
    \item \textbf{Envelope thinning:} The envelope of all the CGM clouds initially has intermediate temperatures ($10^4 <T <10^7\,\rm K$), which make it thermally unstable. Having temperatures corresponding to the peak of the cooling function, all envelopes rapidly cool down, resulting in fast condensation and envelope thinning via isobaric cooling.
    \item \textbf{Compression and inflow driving:} The resulting pressure imbalance between the cooling envelope and the ambient gas then drives inflows towards the centre of the cloud, which leads to core compression and increased densities in the cloud. 
    \item \textbf{Heating and outflow driving:} Cloud compression is halted by radiative heating at distinct volumes (radii) depending on the strength of the heating rates. At the same time, outward pressure waves bounce back from the over-pressurised core, thus creating outflows.
    \item \textbf{Damped pulsations:} When the cloud expands, it once again overshoots pressure balance, which triggers a new compression phase. Rapid compression is halted once more by heating while overshooting pressure balance, and the process repeats over and over again. The process behaves as a damped oscillator with the radiative heating rates controlling the damping rate.
\end{enumerate}

Overall, the panels in Figure \ref{figure:density_maps} and the accompanying movies show that the above stages are universal to all models. The density maps of models standard-cooling, starburst-5kpc and starburst-1kpc reveal a remarkably similar evolutionary path for the morphology of the clouds. This indicates that a $10^4\,\rm K$ cooling floor as well as self-consistent starburst heating rates are all efficient at stopping cooling-driven collapse. The model in panel (b) (which uses the H\&M2012 tables) is an exception as that cloud shrinks considerably more than its counterparts owing to an enhanced compression phase in stage 2. The heating field is simply too weak to halt compression at large radii, resulting in a more compact and denser cloud with a smaller radius than its counterparts.\par

\begin{figure}
    \centering
    \includegraphics[width=0.92\linewidth]{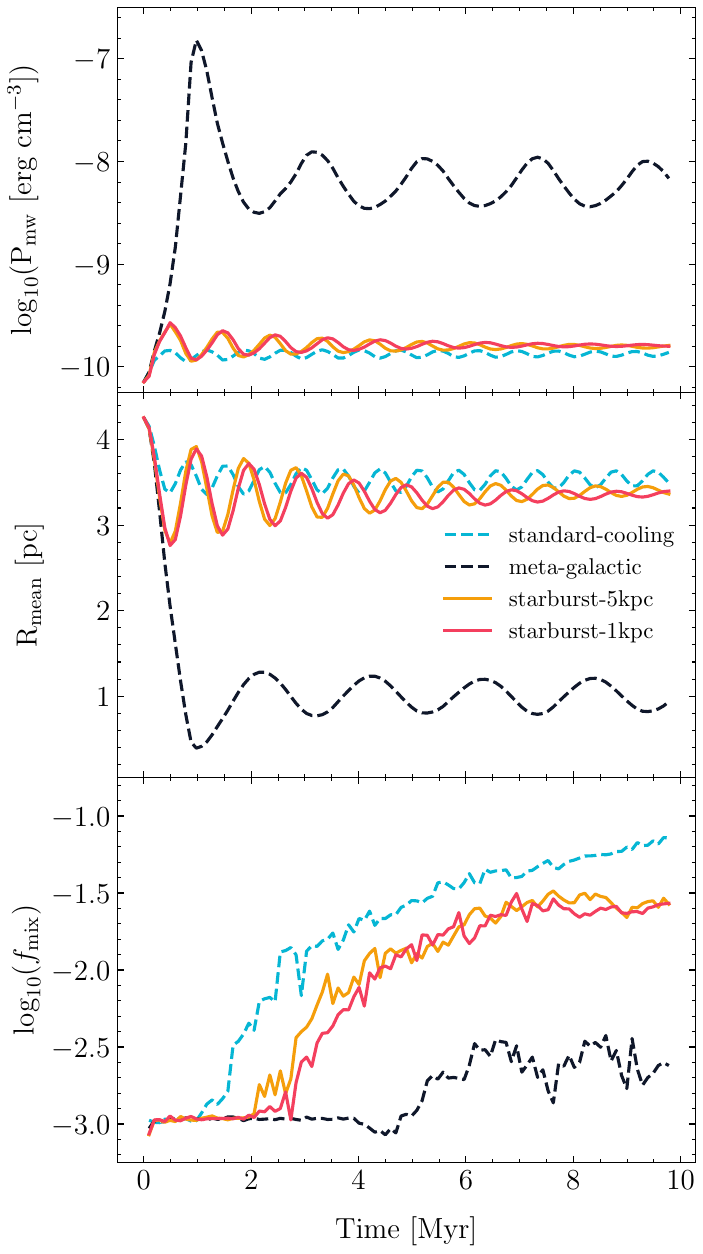}
    \caption{Cloud diagnostics showing the evolution of the mass-weighted thermal pressure (top panel), the mean cloud radius (medium panel), and the mixing fraction (bottom panel) in our 4 models. All models display damped acoustic oscillations regulated by cooling-driven inflows and heating-driven outflows that over-shoot pressure equilibrium. Weak meta-galactic heating/cooling produces more confined clouds; starburst heating/cooling reduce mixing with respect to standard-cooling}
    \label{figure:diagnostics}
\end{figure}

Despite the similitude among models, there are also differences and particular features worth highlighting. To this end, Figure \ref{figure:diagnostics} shows several cloud diagnostics:

\begin{itemize}
    \item The mass-weighted thermal pressure (top panel, $P_{\rm mw}=\int P\,\rho\,C\,dV/\int \rho\,C\,dV$), where $C$ is the cloud Lagrangian tracer, $\rho$ is the mass density, and $P$ the thermal pressure. This diagnostic shows the previously described acoustic oscillations, present in all models. The standard-cooling model settles at a slightly smaller thermal pressure, which in turn corresponds to a slightly larger temperature, than the starburst-5kpc and -1kpc models. In contrast, the meta-galactic model settles at a much larger equilibrium pressure and a much smaller temperature owing to the drastic compression phase this cloud experiences.
    \item The mean cloud radius (medium panel, $R_{\rm mean}=\int r\,\rho\,C\,dV/\int \rho\,C\,dV$), where $r=\sqrt{x^2+y^2+z^2}$. The cloud in the standard-cooling model is slightly larger than its starburst-5kpc and -1kpc counterparts and substantially larger than the meta-galactic cloud, which becomes very confined. This parameter captures the behaviour of the oscillation amplitudes quite well. Unlike the starburst-5kpc and -1kpc models with self-consistent heating that show pronounced damping, the standard-cooling model maintains higher amplitude oscillations. This implies that adding thermal energy self-consistently not only drives outward outflows to counteract cooling-driven inflows and compression, but also has a damping effect on the mechanical oscillations, which seems weaker when a cooling floor is imposed. The meta-galactic cloud model sustains high-amplitude oscillations for a similar reason. Its cooling floor is lower and the heating rates are much weaker, so damping is much slower.
    \item The mixing fraction (bottom panel, $f_{\rm mix}=\int \rho\,C_{\rm mix}\,dV/\int \rho\,C\,dV$), where $C_{\rm mix}=C$ when $0.1\leq C \leq 0.9$. This parameter measures the amount of mixing that occurs between the CGM cloud and the ambient halo material. Mixing in these simulations is driven by dynamical, shear instabilities caused by cooling inflows and heating outflows, and is regulated by the cloud-to-ambient density contrast, $\chi$. The picture revealed by $f_{\rm mix}$ is consistent with that shown by the above diagnostics. The standard-cooling model shows higher mixing than the starburst-5kpc and -1kpc models, owing to the larger amplitude oscillations its cloud is subjected to (despite having the same density contrasts). The heating-driven damping effect reduces shear and mixing by a factor of $3-4$ in starburst models. On the other hand, the meta-galactic cloud model shows very little mixing, $\lesssim 1\,\rm dex$ smaller than the other models. While this cloud is subjected to high-amplitude oscillations, its density contrast $\chi$ is $1\,\rm dex$ larger. Thus, the drastic cooling-driven compression and very high densities inhibit the growth of shear instabilities.
\end{itemize}

\begin{figure*}
    \centering
    \includegraphics[width=\linewidth]{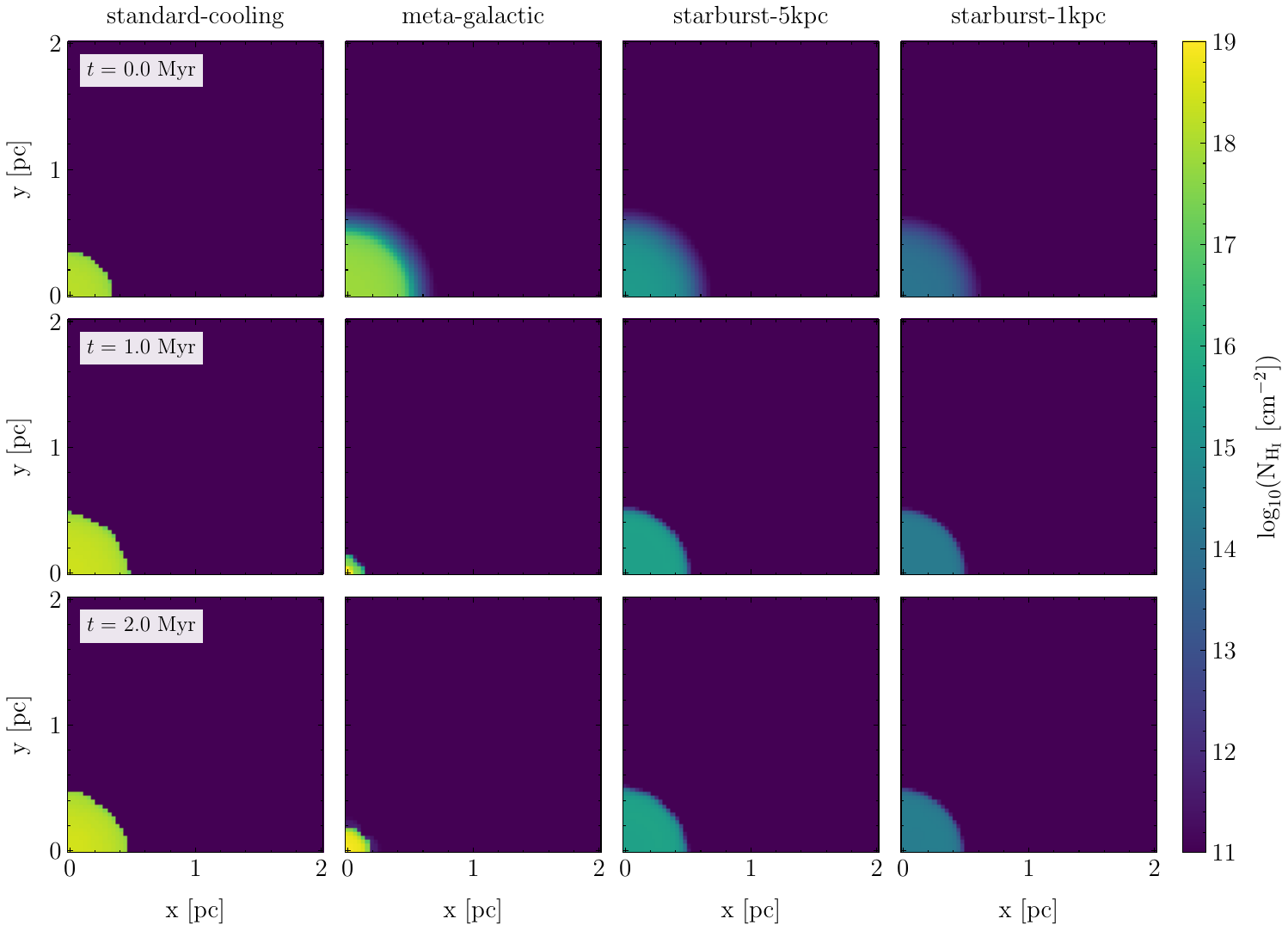}\\
    \caption{Same as figure \ref{figure:density_maps}, but showing the column densities of \text{H\,\textsc{i}}. The standard-cooling model uses a temperature range for \text{H\,\textsc{i}} that overestimates its column densities compared to the starburst models with self-consistent ion fractions. Weak meta-galactic heating produces the highest \text{H\,\textsc{i}} column densities, due to cloud compression. Column densities become lower as the distance to the starburst is reduced, owing to changes in ionisation. Such changes are not captured in standard-cooling approaches.}
    \label{fig:ion_maps}
\end{figure*}

To illustrate the capabilities of \textsc{py4radiation} for post-processing analysis, we present the outputs for a sample ion in our CGM cloud simulations. Figure \ref{fig:ion_maps} shows the time evolution of the neutral hydrogen column density, $N_{\rm HI}$ in our four models (in columns). For the standard-cooling model, we create a mask with characteristic temperature ranges for \text{H\,\textsc{i}} ($10^{2}\leq \log_{10}{\rm T}\leq 10^{4}$) and integrate along the $Z$-axis. For the remaining models, we produce the 2D maps projected onto the $XY$ plane using the \textsc{Trident} code and the \textsc{yt} package, using the ion tables generated with \textsc{py4radiation}. The column number densities span several orders of magnitude, showing significant differences in \text{H\,\textsc{i}} abundances, depending on the adopted ion fraction tables. In general, the spatial distribution of \text{H\,\textsc{i}} follows the morphological evolution described above for all cloud gas, but the column densities differ considerably among models.

The standard-cooling model shows large \text{H\,\textsc{i}} column densities for the CGM cloud, consistently maintaining $10^{18}-10^{19}$ cm$^{-2}$ throughout its evolution. In contrast, the meta-galactic model depicts a spatial contraction of its \text{H\,\textsc{i}} distribution. During the periodic compression phases, the cloud shrinks, confining \text{H\,\textsc{i}} to compact central regions. By $t=2.0\,\rm Myr$, neutral hydrogen is concentrated within a much smaller radius than in the other models. As a result, this model produces the largest \text{H\,\textsc{i}} column densities, following a behaviour consistent with the cooling-driven compression described earlier.\par

The starburst-5kpc and -1kpc models exhibit a similar morphological evolution and spatial distribution of \text{H\,\textsc{i}} to the standard-cooling model, but with considerably smaller \text{H\,\textsc{i}} column densities. In particular, $N_{\rm HI}$ decreases systematically as the cloud is placed closer to the starburst, which reflects the increasing strength of the ionising radiation field. The starburst-5kpc model maintains column densities of $10^{15}-10^{16}$ cm$^{-2}$, while the starburst-1kpc model reaches values closer to $10^{14}-10^{15}$ cm$^{-2}$. We note that in both starburst cases, the column densities are below the typical sensitivity limits of surveys like the Galactic All-Sky Survey (GASS; \citealt{2009ApJS..181..398M}) and the Galactic Australian Square Kilometre Array Pathfinder (GASKAP; \citealt{2022PASA...39....5P}), which map atomic hydrogen clouds in and near our Galaxy.\par

Overall, Figure \ref{fig:ion_maps} illustrates the importance of accounting for self-consistent ion fractions in simulation post-processing analyses, which \textsc{py4radiation} allows. While the standard-cooling and starburst models follow similar evolutionary paths in terms of their density distributions and mean radii, their \text{H\,\textsc{i}} column densities differ by several dex when utilising self-consistent ion tables. Using temperature ranges to isolate \text{H\,\textsc{i}} overestimates column densities. On the other hand, the meta-galactic model shows how weak radiative heating can modify the neutral hydrogen distribution by allowing stronger compression and the formation of a compact, dense core.

\section{Conclusions}
\label{section:conclusions}
We have reported \textsc{py4radiation}, which is a novel, publicly-available, Python-based, open-source, and parallelised package that acts as an integration framework around the \textsc{Cloudy} code to produce customised heating/cooling functions and ion fractions, all consistent with input ionising UV backgrounds. Our package takes as inputs a meta-galactic background radiation field from \cite{HaardtMadau2012} or a custom starburst spectral energy distribution (SED) from \textsc{Starburst99} (or other packages). For starburst SEDs, it appropriately normalises and scales them to a flux density at a desired distance from the source. Then, it structures the input files, so they match the format required by the spectral synthesis \textsc{Cloudy} code, and subsequently runs \textsc{Cloudy} for a user-defined $(n_{\rm H},T)$ matrix that represents the desired gas conditions. The user can easily adjust the initial $(n_{\rm H},T)$ parameters, as well as the run mode using a configuration file. The numerical execution is handled by \textsc{pyCIAO}, whose multi-dimensional output arrays are then formatted into tables: in \texttt{ASCII} format for heating and cooling functions and in \texttt{HDF5} format for ion fractions. The \textsc{py4radiation} integration framework can also be set to run \textsc{Cloudy} in parallel on CPU architectures by using either the \texttt{multiprocessing} or the \texttt{MPI} library. Our scaling study shows that the \textsc{py4radiation} parallelisation is robust, and allows users to significantly speed up code execution and output retrieval.\par

Radiative heating, cooling, and photo-ionisation play an important role in determining the thermal and dynamical evolution of astrophysical gases, particularly in environments where multiphase structures are exposed to ionising radiation from nearby stellar populations or the meta-galactic background. The development of \textsc{py4radiation} was motivated by the lack of an astrophysical tool with detailed documentation that can easily and rapidly produce radiative cooling and heating functions alongside ion tables from custom input SEDs. While \textsc{Cloudy} and the Perl-based \textsc{CIAOLoop} provide these capabilities, the user must manually handle the inputs and configuration. \textsc{py4radiation} provides a modern and parallelised architecture that substantially eases SED preparation, wraps the code execution (without abstracting the user from important configuration options), and natively handles output formatting. It does so in a way that the outputs can be readily coupled to existing hydrodynamic codes (e.g. \textsc{PLUTO}/\textsc{gPLUTO}) and spectral post-processing packages (e.g. \textsc{Trident}). Our package provides the community with a fast and robust tool to simulate the effects of radiation fields on astrophysical gases on the optically-thin limit, as we demonstrate with a sample simulation targeting a static CGM cloud embedded in a hotter ambient medium. Our simulation suite studies the thermodynamic history of an idealised, static CGM cloud with smoothed edges subjected to a meta-galactic UV/X-ray background and two starburst UV backgrounds placed at different distances from the radiation source. \textsc{py4radiation} handles the input SED and provides self-consistent heating/cooling functions, which we subsequently couple to our hydrodynamical \textsc{gPLUTO} simulations. At the post-processing stage, we use the ion tables, also produced by \textsc{py4radiation}, to study the evolution of the \text{H\,\textsc{i}} ion.\par 

Our CGM cloud simulations indicate that the cooling floor of $10^4\,\rm K$, widely used in the simulation literature, halts cooling-driven compression in a reasonable fashion (for our initial conditions) compared to self-consistent models with starburst heating. Despite that, the standard-cooling approach produces a slightly more enlarged cloud that sits at a lower equilibrium pressure (corresponding to a somewhat higher temperature than that of its self-consistent counterparts). The $10^4\,\rm K$ cooling floor also fails to capture the damping effect produced by thermal energy injected via starburst heating, which artificially increases mixing fractions by factors of $4$-$5$ in that model. Similarly, taking a temperature range to isolate \text{H\,\textsc{i}} in the standard-cooling model overestimates its column densities by several dex. Thus, our results suggest that including heating and ion populations self-consistently does have an effect on mixing processes and ion abundance predictions, which can be important in simulations targeting ISM and CGM conditions as well as in their comparisons to observations. The cloud model subjected to meta-galactic heating exhibits a unique behaviour in our suite. Heating rates are weak and unable to halt cloud compression or dampen the strong pulsations driven by flipping pressure gradients. The cloud in that scenario acquires a very large density contrast, which makes it less prone to shear instabilities and mixing. Further work into heating effects for a broader set of initial conditions is certainly warranted.\par

While our results demonstrate the applicability of \textsc{py4radiation} to modelling the effects of radiation fields on astrophysical gases, some limitations should be acknowledged. The heating/cooling functions presented here are computed in the optically thin limit, which is an approximation commonly adopted in tabulated cooling prescriptions. Therefore, our cloud simulations do not account for self-shielding or the attenuation of ionising radiation within dense gas, which could affect the local thermal structure and ionisation states. Future development of \textsc{py4radiation} will include updating the underlying version of \textsc{Cloudy} to more recent releases and extending the range of supported stellar population synthesis models beyond \textsc{Starburst99}. These improvements will expand the applicability of the package to different astrophysical environments and radiation fields.

\section*{Software and Data availability}
The \textsc{py4radiation} source code is publicly available at \href{https://github.com/cPhysPlus/py4radiation}{Github Code Repository}, and is distributed under the MIT licence with documentation at \href{https://py4radiation.readthedocs.io}{ReadTheDocs}. The heating/cooling functions and ion tables generated with the software are publicly available through the project data catalogue at \href{https://github.com/cPhysPlus/heating-cooling-ion-tables}{GitHub Dataset}. The datasets are released under the CC BY 4.0 licence.

\section*{Acknowledgements}
This work was carried out as part of research project PHY23-03 at Universidad Yachay Tech, Ecuador. The authors gratefully acknowledge the Gauss Centre for Supercomputing e.V. (\url{www.gauss-centre.eu}) for funding this project by providing computing time (via grant pn34qu) on the GCS Supercomputer SuperMUC-NG at the Leibniz Supercomputing Centre (\url{www.lrz.de}). In addition, the authors thank CEDIA (\url{www.cedia.edu.ec}) for providing access to their HPC cluster as well as for their technical support. This work has made use of the VISIT visualisation software (\citealt{HPV:VisIt}). We also thank the developers of the \textsc{PLUTO}/\textsc{gPLUTO} codes for making these hydrodynamic codes available to the community. BC acknowledges support from Cannon Fellowship.

\bibliographystyle{apsrev4-1}
\bibliography{py4radiation}

\begin{appendix}

\section{Comparison with other cooling curves}
\label{ap:comparison}

\begin{figure}[!ht]
    \centering
    \includegraphics[width=0.92\linewidth]{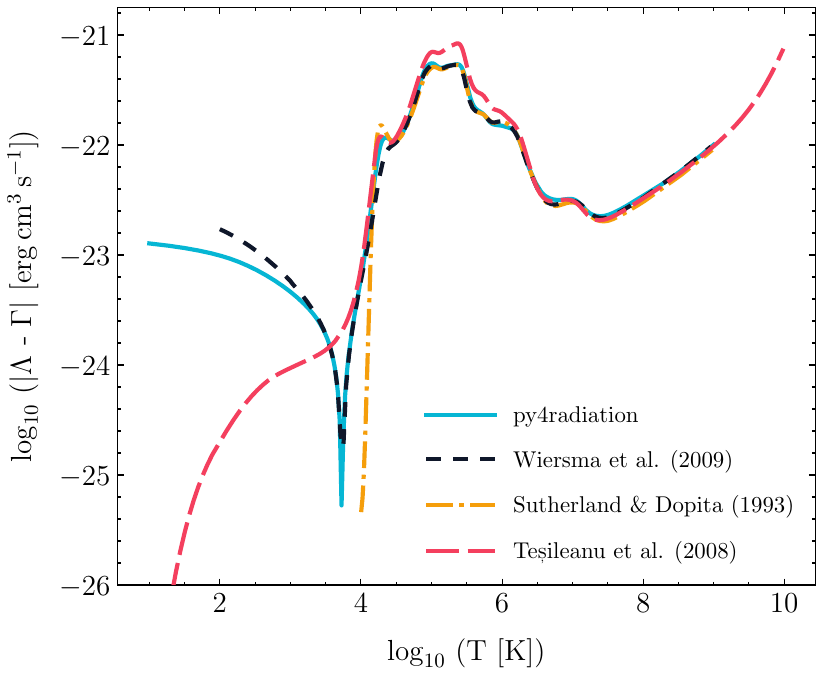}
    \caption{Comparison of the net heating/cooling functions produced by \textsc{py4radiation} (solid) for $n_{\rm H}=0.01$ ${\rm cm}^{-3}$ with functions by \cite{Sutherland1993} (dash-dotted), \cite{Tesileanu2008} (long-dashed), and \cite{Wiersma2009} (short-dashed). Overall, the results are consistent with established calculations.}
    \label{fig:cooling_comparison}
\end{figure}

Here we compare the net heating/cooling function produced by \textsc{py4radiation} for $n_{\rm H}=0.01$ ${\rm cm}^{-3}$ using the redshift-zero, meta-galactic radiation field with other heating/cooling functions from the literature by \cite{Sutherland1993}, \cite{Wiersma2009}, and \cite{Tesileanu2008} (see Figure \ref{fig:cooling_comparison}).\par

Overall, the comparison demonstrates that \textsc{py4radiation} produces tabulated heating/cooling rates consistent with well-established calculations. There are also some differences, which can be attributed to the adopted physical assumptions and the treatment of radiative heating/cooling. For instance, the \cite{Sutherland1993} curve has a temperature floor at $10^4\,{\rm K}$, \cite{Wiersma2009} adopted the older \cite{HaardtMadau2001} background, and \cite{Tesileanu2008} used an earlier version of \textsc{Cloudy} 90.01. The comparison with \cite{Wiersma2009} is the cleanest as they also report heating/cooling rates, which are in overall agreement with ours.

\end{appendix}

\end{document}